\documentclass[twocolumn,showpacs,preprintnumbers]{revtex4}
\usepackage{graphicx}

\begin{document}

\preprint{APS/123-QED}
\title{Wet sand flows better than dry sand}
\author{Jorge E. Fiscina}
\author{Christian Wagner}
\email{c.wagner@mx.uni-saarland.de}
\affiliation{Technische Physik, Saarland University, D-66123, Saarbr\"{u}cken Germany }
\date{\today }

\begin{abstract}
We investigated the yield stress and the apparent viscosity of sand with and
without small amounts of liquid. By pushing the sand through a tube with an
enforced Poiseuille like profile we minimize the effect of avalanches and
shear localization. We find that the system starts to flow when a critical
shear of the order of one particle diameter is exceeded. In contrast to
common believe, we observe that the resistance against the flow of wet sand
is much smaller than that of dry sand. For the dissipative flow we propose a
non-equilibrium state equation for granular fluids.
\end{abstract}

\pacs{05.70.Ln, 83.60.La, 45.70.Mg, 83.80.Fg.}
\maketitle

Capillary forces between single grains are the cause of the stiffness of
sculptured wet sand, as opposed to dry sand which does not support its own
weight \cite{Horn}. Therefore, it is commonly believed that wet sand should
be more resistant, i.e. more "viscous", than dry sand. Not many studies,
though, on the flow behavior of wet sand have been undertaken yet. The few
studies at hand have been performed mostly either in a rotating drum \cite%
{Gollub} or in standard rheometric set-ups \cite{Bonn06}. Our study was
inspired by an apparatus invented by the authors of references \cite{Ge04,
Fo05, He05}, but their work is more focused on the regime below the yield
point. Besides its practical applications in e.g. civil engineering or
industrial processes, the rheological or dynamical properties of dense
packed granular materials are investigated to help develop a non-equilibrium
thermodynamic formalism. Whilst the description of a rapid flowing granular
gas is well established \cite{Bagnold}, the opposite limit of a dense
packing remains an open question. Proposals for state equations for granular
matter include, for example, the Enskog equation either for describing rapid
dilute granular flows \cite{Ernst} or for the high density limit \cite{Hong}
. Yet, it is not possible to describe all of the experimental findings, e.g.
the fact that granular matter does not follow a Boltzmann statistic \cite%
{HHo97, PRL9510}.

Granular matter is an example of a so-called yield stress fluid. Above a
critical yield stress, the mechanical response changes from a solid-like or
jammed state to a flowing one. An illustrative study on the rheological
properties of wet soils was given, for example, in ref. \cite{Gh1}. These
authors measured viscoelastic properties under steady and oscillatory
stress. Soils are a typical example of wet granular material, and they
exhibit a mixed behavior of elasticity, viscosity and plasticity that is
mainly determined by the quantity of water, the yield stress, shear modulus,
and viscosity decrease with the water content. The water content is changed
by adding liquid to a given sample, thus reducing the packing density of the
solid grains. The system starts to flow when the external stress exceeds the
inter-aggregate contact \cite{Da52} forces. The mechanical properties at
lower water content are determined by the liquid bridges between clay
clusters and quasi-crystals, and those at higher water content are
determined by the flow of the liquid through the soil pores \cite{Gh1}.

A principle difficulty in the determination of the yield stress
point arises from the fact that a yield stress fluids exhibits
thixotropy, also referred to as aging or rejuvenation
\cite{Bonn06}. Yield stress and thixotropic behavior depend on the
degree of compaction and the topology. Both can be modulated by
weak attractive forces like the electrostatic charge and humidity
\cite{Rhodes}. Macroscopically, the jammed nature of dry and wet
granular materials determines the viscoelastic properties, e.g.
the route to the jammed transition can be explained by a modified
Vogel-Fulcher-Tammann behavior \cite{DA01, DA03}, and it can be
understood by the relation between the compaction and the grain
mobility \cite{DeGennes,Lumay}. In dense granular materials the
resistance to flow and the dissipation of energy occurs when the
bulk granulate has to pass from one configuration to another. This
rearrangement process leads to a new, jammed configuration. The
rearrangement dynamics driven by shearing causes the rejuvenation
and aging of the fluid-structure and it is meaningful to determine
the yield stress point when the structural integrity is maintained
and shear localization, avalanches, and arching effects are
prevented. We will show that by doing so, it is possible to reach
a steady state with a constant relation between the pumping and
dissipation of energy if the granular material is sheared in a
oscillatory way.
\begin{figure}[tbp]
\includegraphics[width=1\linewidth,angle=0]{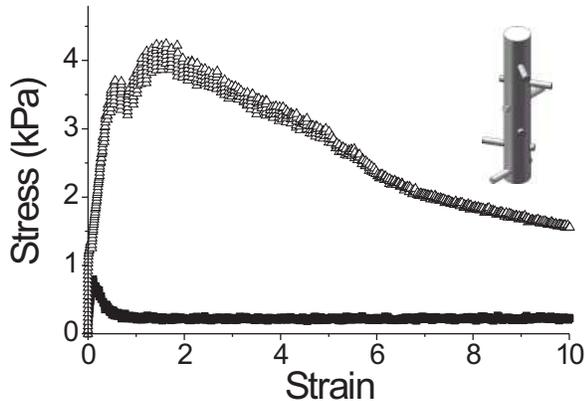}
\caption{Comparative rheological test under constant deformation
(CD modus) for micron-sized dry sand (filled black squares) and
the wet paste with a water content of $W=0.05$ (open triangles).
In the inset, the sensor for the "bars" geometry is shown.}
\label{fig1}
\end{figure}
Our sand was composed of glass spheres of diameter $D=140-150\mu$m with and
without additional deionized water. The content of water in the paste $W$ is
defined as the quotient between the liquid volume and the volume occupied by
the grains. As a first attempt to measure the yield stress point of our
samples, we followed the methods used to study the mechanics of soil.
Measurements were done with a rotational viscometer (MARS, Thermo Haake,
Karlsruhe, Germany) with a parallel plate geometry \cite{Gh1}, and with sand
paper glued on the on the surfaces of the plates. Further improvement was
given by using a vane sensor or a "bars" sensor, which is shown in the inset
of fig. \ref{fig1}. However, for our study, all geometries gave comparable
results, while the latter was the most reproducible. A typical protocol to
study yield stress fluids - for example coatings and paints - involves
imposing a constant shear rate and measuring the strain. The so-called yield
stress point is the highest stress at which no flow occurs at a given shear
rate, e.g., at the maximum in the flow curve of fig. \ref{fig1}. Beyond the
maximum, the deformation is large enough to compromise the structural
integrity of the granular fluid. Figure 1 shows a representative result for
the case studies of dry and wet sand. The maximum corresponds to the stress
at which the material begins to plastically deform; before that, the
material deforms elastically. The test must be carefully tuned, because the
shear rate must be small enough to catch the elastic and plastic regimes
but, if it is too small, the peak is difficult to identify. In general, the
yield stress point depends on the time scales and shear rates applied in the
experiment - however, it was recently found that the critical yield strain
of many different yield stress fluids, like ketchup or pastes, does not
depend on the external parameters; instead, it depends only on the type of
fluid \cite{Caton}.

The yield stress point for the dry sand is $\tau =0.8k$Pa, $\gamma
_{c}=0.1=D/10$ ($\tau$ is the stress and $\gamma$ the strain)
while, for the wet sand, $\tau =4.23k$Pa, $\gamma _{c}=1.4=3D/2$,
for a given shear rate of $\dot{\gamma}=0.0038$ s$^{-1}$ and
$\dot{\gamma}=0.003$ s$^{-1}$ respectively. Apparently, the liquid
bridges between the grains reinforce the structural integrity of
the wet sand. Similarly, one needs less torque to turn the sensor
at a given rotational speed in the dry sand than the wet sand. The
asymptotic values of the stress are $0.2$ and $1k$Pa respectively;
however, due to the breakdown of the microstructure and flow
localization, one cannot straightforwardly deduce the yield stress
(or even a shear viscosity) later on.
\begin{figure}[tbp]
\includegraphics[width=1\linewidth,angle=0]{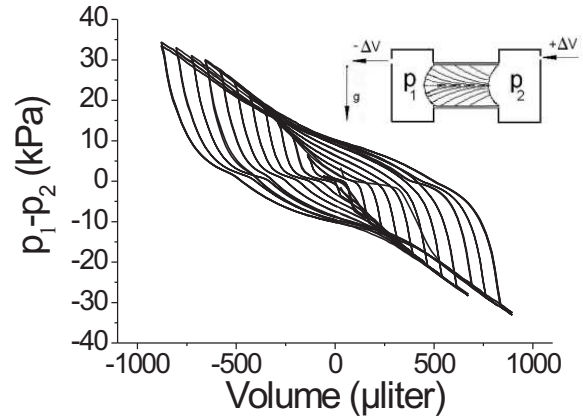}
\caption{Family of differential pressure curves for increasing shear
amplitudes (see text for further explanation); the inset shows a schema of
the measurement cell.}
\label{fig2}
\end{figure}
To override this problem, we carried out experiments with the shear cell
proposed in ref.\cite{Ge04, Fo05, He05}(see the inset of fig. \ref{fig2}).
The use of this device made it possible to shear the pastes all over the
sample volume. The granular material is put into a acrylic glass cylindrical
cell at atmospheric pressure. The sides of the cell are sealed with thin
latex membranes of $300{\mu}$m, which are flat at the beginning of the
experiment. Chambers filled with water are attached to both sides of the
cell, adjacent to the membranes, where it is possible to inject and extract
water into and out of the chamber through tubes via a syringe. The
syringe-pistons are connected to spindles that could be moved with a stepper
motor. The pistons move at the same speed as each other but in the opposite
direction to prevent dilatation. The acrylic glass tube we used for this
experiment was $24$ mm long. It was filled with dry sand or with a wet sand
paste with $W=0.03$ so that, in both cases, the global packing fraction $%
p_{f}=0.63$. We should mention that this number must be considered
cautiously, since we ignore the local compaction of the granulate, e.g., the
topology of polytetrahedral grain structures \cite{An07}.

The deformation of the membranes is approximately parabolic and imposes a
Poiseuille-like flow profile within the sample; thus, avalanching and
arching can be kept to a minimum. Piezoresistive sensors sense the pressure
(with respect to atmospheric pressure) in both chambers. A fixed quantity of
water $\Delta V$ is injected or extracted in/from the adjacent chambers, and
the differential pressure $p_{1}-p_{2}$ is measured (see the inset of fig. %
\ref{fig2}). The displacement of the membrane is calculated as $\Delta
x=4\Delta V/\left( \pi D_{c}^{2}\right) $, where $D_{c}$ is the diameter of
the acrylic glass cell ($24m$m). The paste is sheared at a very low shear
rate with a shear amplitude $a_{s}$, between $-a_{s}$ and $a_{s}$. The
protocol was to move the membrane $\Delta x=6{\mu }m$ in one second, wait $13
$ seconds, and measure $p_{1}$ and $p_{2}$ until the selected $a_{s}$ is
reached. A curve of the pressure difference $p_{1}-p_{2}$ as a function of
the displacement - which represents the strain - is obtained for each
shearing amplitude $a_{s}$. A family of differential pressure
characteristics for increasing shear amplitudes are shown in fig. \ref{fig2}%
. The slope of the two branches corresponds to the restoring force
exerted by the rubber membranes, but a stress $\tau _{st}$ is
stored in the system which is related to the opening of the loop
with $2\tau _{st}(a_{s})=\Delta (p_{1}-p_{2})_{0}$.
\begin{figure}[tbp]
\includegraphics[width=1\linewidth,angle=0]{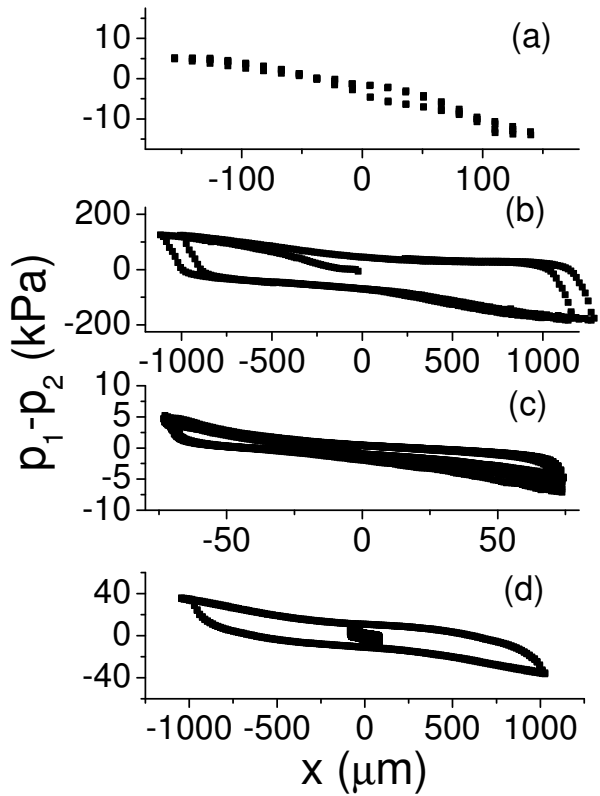}
\caption{Differential pressure versus the shear displacement. (a),(b): A
sample of 17.7 gr. of dry sand ($26\%$ room humidity,$26 C^\circ$) is
compared with (c),(d): a paste of the same sand content mixed with water ($%
W=0.03$).}
\label{fig3}
\end{figure}
\begin{figure}[tbp]
\includegraphics[width=1\linewidth,angle=0]{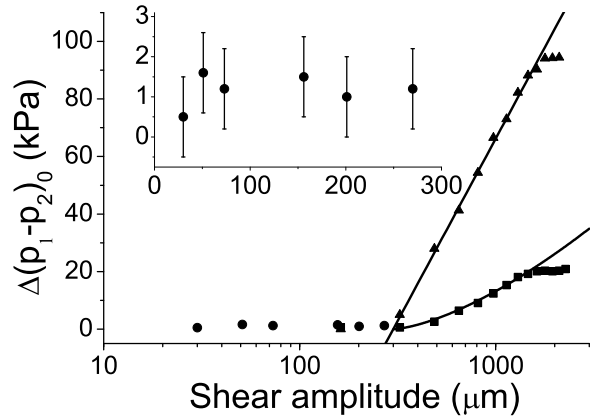}
\caption{Stress - strain behavior for increasing shear amplitude for dry
(triangles) and wet sand $W=0.03$ (squares) and the corresponding fittings
for the dissipative flow range (equation (\protect\ref{Magic})). The inset
shows, for the wet sand, the range where the stored stress is the yield
stress.}
\label{fig4}
\end{figure}
In fig. \ref{fig3}, the differential pressure curves for dry and wet sand
are shown for the two limiting cases of small and large shear amplitudes. By
shearing with $a_{s}<$ $2D$, we find that the dry sand does not resist any
stress (fig. \ref{fig3} (a)), while the wet sand behaves like a yield stress
fluid due to the liquid bridge network (LBN) (fig. \ref{fig3} (c) and the
inset of fig. \ref{fig4}), with a $\Delta (p_{1}-p_{2})_{0}\approx $ $1k$Pa.
Following ref.\cite{Ge04, Fo05, He05}, the capillary pressure is an energy
density $p=dE/dV\leq p_{f}N\gamma /D$ where $N$ $=6$ is the LBN\
coordination number, $\gamma $ is the surface tension of water, and $%
D=145\mu $m; thus, $p\leq 1.9k$Pa and, in agreement with ref.\cite{Ge04,
Fo05, He05}, we find that the restoring forces originate from the capillary
forces between the grains. The onset of dissipative flow occurs at $%
a_{s}\geq $ $2D$ then the opening of the loop starts to increase. It grows
more strongly for the dry than for the wet sand, with maximum values of $116k
$Pa, and $22k$Pa for the largest shear amplitude, respectively. This means
that one needs \textit{less energy to push the wet sand through the tube}.
It is worth mentioning that, for a given shear amplitude, the hysteresis
loop is stable during a many cycles. We carried out a few tests which were
one week in duration and found that the opening of the loop for different
shear amplitudes remains constant (we made sure that all our data were taken
in such a steady state situation). We also tested shear velocities between $%
0.6$ and $300{\mu }$m$/$s without observing differences in the hysteresis
loop.

Figure \ref{fig4} shows the complete \textit{stress-strain behavior} for the
dry and the wet sand. The inset shows the differential pressure for wet sand
for $a_{s}<$ $2D$, from which we deduce a mean value for the yield stress $%
\tau =\left\langle \Delta (p_{1}-p_{2})_{0}\right\rangle /2=\left( 0.6\pm
0.5\right) k$Pa. For the dry sand, we find (in this range) $\left\langle
\Delta (p_{1}-p_{2})_{0}\right\rangle =\left( 0\pm 1\right) k$Pa. In the
dissipative flow range $a_{s}\geq 2D$, it is not possible to shear the
sample without changing the local compaction by rearrangement of the grains,
and the response of the system is a change of configurations between jammed
states. The experiment shows that the addition of a small amount of water
changes the flow resistance dramatically, possibly due to the existence of
lubrication layers and because it favors cluster formation that minimizes
the effective free volume\cite{An07}. Near the onset of yield, the
stress-strain curve for increasing $a_{s}$ (fig. \ref{fig4}) can be modeled
with a stretched exponential:
\begin{equation}
\frac{a_{s}}{2D^{\ast }}=\exp \left( \frac{\Delta (p_{1}-p_{2})_{0}}{c_{0}}%
\right) ^{n}  \label{Magic}
\end{equation}%
A fit with with equation (\ref{Magic}) yields an onset value $D^{\ast
}=150\mu $m for both cases which is close to the mean diameter of the
grains; for the constant $c_{0}$ and the exponent $n$, we find $c_{0}$ $=55$%
, $n=1 $ and $c_{0}=10k$Pa, $n=3/2$, for the dry and wet sand respectively.
The work that is applied to the system is given by $W=F.a_{s}$, and its
response is related to the energy $E$ that is dissipated during the
rearrangement mechanism; thus, it is possible to rewrite eq. (\ref{Magic}) as

\begin{equation}
\frac{W}{W_{0}}=\exp \left( \frac{E}{E_{0}}\right) ^{n}
\label{State
Equation}
\end{equation}

where $W_{0}=2F.D^{\ast }$ and $E_{0}$ is the activation energy for the
rearrangement. The latter should be related to a configurational entropy; we
find $E_{0}=88$ and $16n$J per grain for the dry and wet granulate
respectively.

For $n=1$, eq. (\ref{Magic}) it is a Doolittle-like equation \cite{Doo}. It
was shown that the viscosity of a granular fluid depends strongly on the
particle concentration \cite{Huang} and, in terms of a free volume theory, $%
E\propto 1/v_{f}$. The activation energy is related to the global packing
factor $E_{0}\propto p_{f}$ (the occupied volume divided by the cell
volume). Following emerging concepts on non-equilibrium thermodynamics \cite%
{Ritort}, we propose that the equation (\ref{State Equation}) could be a
state equation for granular fluids in dissipative flow under globally
constant packing fraction conditions.

It is difficult to estimate the value of $E_{0}$ for dry sand since the
energy loss is sensitive to, e.g., microscopic roughness and topology. But,
for the wet sand, it can be related to the liquid bridges. By considering
the coordination number $N=6$, the dissipation per bridge is $E_{0}/N=$ $%
(2.6\pm 0.5)n$J, which is in good agreement with the estimation
for the energy loss during bridge formation and rupture $\Delta
E_{cap}<\pi \gamma D^{2}/2=2.4n$J \cite{He05}.
\begin{figure}[tbp]
\includegraphics[width=1\linewidth,angle=0]{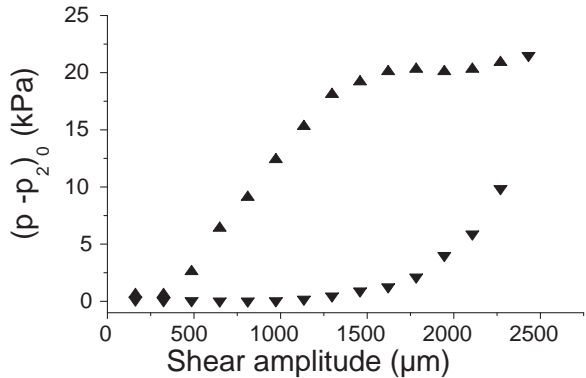}
\caption{Stress - strain curve for wet sand obtained by increasing
(triangles up) and decreasing the shear amplitude (triangles down).}
\label{fig5}
\end{figure}
Figure \ref{fig5} shows a stress-strain curve for the first scan of
increasing and decreasing the shear amplitude applied to wet sand. Beyond $%
a_{s}$ $\geq 2D$, the structure of the granular fluid changes during the
successive rearrangements. Possible mechanisms for this include the
formation of water channels by coalescence of the liquid bridges and larger
polytetreahedral solid structures. The decreasing branch of Fig.5 might thus
correspond to a fluid that exhibits phase separation and shear bands. For
the range $a_{s}<2D$, the opening of the loop (the yield stress) tends to
zero; again, this is an indication for a phase separation. Similar
measurements for the dry sand revealed no difference between increasing and
decreasing shear amplitude, but the rearrangement process there occurs not
between evolving clusters, but between the single grains.

In conclusion, we have found that it is much easier to push wet sand than
dry granular matter in a Poiseuille-like profile through a tube. Even if the
capillary forces increase the yield stresses, it looks like the water
promotes the cluster formation and reduces the inter-grain friction. The
leading dissipation mechanism results from the rupture and formation of
liquid bridges, and we are able to explain our data quantitatively within
the framework of an excluded volume theory. Finally, we find indications
that the yield of the system is related to the microscopic size of the
grains. Further studies with different grain sizes are ongoing.

\textit{Acknowledgments}. The work was funded by the Alexander von Humboldt
Foundation and the DFG-Graduiertenkolleg 1276/1. We thank Manuel C\'{a}ceres
and Nicolas Vandewalle for enlightened discussions.

\end{document}